\documentclass[journal]{IEEEtran} 				
\IEEEoverridecommandlockouts 						
\usepackage{amsmath,amssymb}
\usepackage{amsthm}
\usepackage{color}
\usepackage[mathscr]{eucal}
\usepackage{graphics,graphicx,multicol}
\usepackage{epsfig}
\usepackage{enumerate}
\usepackage{subfigure}
\usepackage{algorithmic}
\usepackage{algorithm}
\usepackage{morefloats}
\usepackage{enumerate}
\usepackage{subfigure}
\usepackage{multirow}
\usepackage{array}
\usepackage{pstricks, pst-node, pst-plot, pst-circ}
\usepackage{moredefs}
\usepackage{refstyle}

\usepackage{courier}

\author{Zaid Kbah}
\begin{document}
\title{Resource Allocation in Cellular Systems for Applications with Random Parameters} 
\author{\IEEEauthorblockN{Zaid Kbah\IEEEauthorrefmark{1}, Ahmed Abdelhadi\IEEEauthorrefmark{2}}\\
     \IEEEauthorblockA{\IEEEauthorrefmark{1}University of Technology, zkbah1@unh.newhaven.edu},\\
     \IEEEauthorblockA{\IEEEauthorrefmark{2}Hume Center, Virginia Tech, aabdelhadi@vt.edu
     } 
     }
\maketitle

\begin{abstract}
In this paper, we conduct a study to optimize resource allocation for adaptive real-time and delay-tolerant applications in cellular systems. To represent the user applications via several devices and equipment, sigmoidal-like and logarithm utility functions are used. A fairness proportional utility functions policy approach is used to allocate the resources among the user equipment (UE)s in a utility percentage form and to ensure a minimum level of customer satisfaction for all the subscribers. Meanwhile, the priority of resources allocation is given to the real-time applications. We measure the effect of the stochastic variations of the adaptive real-time applications on the optimal rate allocation process and compare the results of deterministic and stochastic systems. Our work is distinct from the other resource allocation optimization works in that we use bisection method besides convex optimization techniques to optimally allocate rates, and we present the adaptive real-time applications in a stochastic form. We use Microsoft Visual Basic for Applications with Arena Simulation Software interface to simulate the optimization problem. Finally, we present our optimization algorithm results. 
\end{abstract}

 
\section{Introduction}\label{sec:intro}
Since the issue of seminal network utility maximization presented in \cite{1}, interest in optimizing resources allocation has radically increased for a variety of fields. In \cite{1}, the authors address the issues of sharing the available bandwidth optimally between user applications through a two-stage rate control algorithm. The authors present a mathematical model to analyze the stability and fairness issues, which generally are the main concern of engineers and economists, of the rate control algorithm. Kelly presents a rate control algorithm based on shadow price of one bandwidth unit to achieve proportional fairness. In \cite{2}, the authors present an algorithm to maximize the total UEs rate. The dual problems at the source, UEs, and network (eNodeB) were solved through asynchronous distributed algorithm. A decentralized scheme was used to execute and connect the results of the local algorithms. Considering the work that was done in both \cite{1} and \cite{2}, the logarithm utility function that was used, which has the best representation to the network traffic for wired communication network, is a concave function. With the increasing demand on multimedia and wireless adaptive real-time applications, another utility function that is non-concave was introduced to address this issue in \cite{3}. Sigmoidal-like utility functions were used to represent adaptive real-time applications, such as video streaming. In this paper, we introduce a study that we conducted to find the optimal resource allocation among UEs in cellular networks. We use the convex optimization techniques with bisection method to allocate the resources optimally and to obtain a minimum quality of service (QoS) among users, while prioritizing adaptive real-time applications. Stochastic programming is a very effective tool to precisely model the variability in UEs rate. Trying to represent UEs demand by collecting a huge amount of data and monitoring the historical trends could be time consuming and expensive as well. As a consequence, the stochastic programming has been used to overcome unnecessary effort and cost. A major issue in allocating resources among users is optimally allocate these resources and give the priority to the adaptive real-time users, whose applications are not tolerant to any delays in services, while maintaining a minimum QoS to the delay-tolerant applications.
\subsection{Related Work}\label{sec:related}
In \cite{4}, the authors present a distributed power allocation algorithm based on dynamic pricing of the sigmoidal-like utility functions for a mobile cellular network. The two-stage power allocation algorithm approximates the optimal allocation of resources, but fails to maintain a minimum QoS. In \cite{5}, the author presents an algorithm to solve the utility function of UEs with multiple applications in a cellular network. The proposed algorithm is a modified version of the rate control algorithm in \cite{1} that solves elastic and inelastic utility functions. The relationship between optimal resource allocation and carrier aggregation discussed in \cite{5}. In \cite{6}, the authors present a distributed algorithm to optimally allocate resources, from multiple carriers, for users with logarithmic and sigmoidal-like utility functions applications while maintaining a minimum QoS. In \cite{7}, the authors propose a solution to optimally allocate resources for logarithmic and sigmoidal-like utility functions applications in wired and wireless networks. In \cite{8}, the author presents utility max-min fairness to optimally allocate resources for real-time and delay-tolerant users who are sharing a single path in the system. In \cite{9}, the authors present a distributed rate algorithm that gives priority to the real-time applications and considers a minimum QoS to the other applications in the wireless network. 
\subsection{Our Contribution}\label{sec:our contribution}
Our contribution is summarized below:
\begin{itemize}
\item We solve the utility functions of both convex (sigmoidal-like) and strictly concave by using convex optimization techniques and bisection method. In our optimization problem, priority is given to the adaptive real-time applications while allocating a minimum QoS to the other applications (time-tolerant applications).
\item We assume that UEs utility functions are continuously changing within a certain range. We incorporate uncertainty to the sigmoidal-like utility functions parameters, by using stochastic programming, to increase the robustness of the system while satisfying the mission of the resource allocation optimization problem. Normal and triangular distributions have been assigned to obtain accurate results.
\end{itemize}
The structure of this paper is organized as follows. Section \ref{sec:PROBLEM FORMULATION} presents the optimization problem formulation. In Section \ref{sec:DISTRIBUTED OPTIMIZATION ALGORITHM}, we present our distributed optimization algorithm. In Section \ref{sec:Rate Variations}, we discuss the variations of rate in the adaptive real-time applications and stochastic modeling. In Section \ref{sec:SIMULATION RESULTS }, we discuss our simulation results. Finally, Section \ref{sec:CONCLUSION }, concludes our paper.

\section{PROBLEM FORMULATION }\label{sec:PROBLEM FORMULATION}
In our paper, we consider a single cellular network with $M$ number of users and a base station. The allocated rate by the base station is to the $i^{th}$ UEs and is given by $r_i$ where $i=({1,2,3,....M})$. The UEs are represented by utility functions $U_i(r_i)$. UEs utility functions differ as the UE run different applications. We assume that some of the UEs have utility functions that are strictly concave (logarithmic utility functions), while others have convex utility functions (sigmoidal-like utility functions). Our objective is to distribute the resources, at the base station, optimally among UEs. The utility functions are characterized as follows: 
\begin{itemize}
\item $U_i (0) = 0$
\item $U_i (\infty) = 1$
\end{itemize}
In our paper, we use sigmoidal-like utility functions to represent UEs, as in \cite{4} and \cite{15}. It can be written as
\begin{equation}\label{eqn:sigmoidal}
U_i(r_i) = c_i\Big(\frac{1}{1+e^{-a_i(r_i-b_i)}}-d_i\Big)
\end{equation}
where $c_i = \frac{1+e^{a_ib_i}}{e^{a_ib_i}}$ and $d_i=\frac{1}{1+e^{a_i b_i}}$. The sigmoidal-like utility functions satisfy $U_i(0)=0$ and $U_i (\infty)=1$. In addition, we use logarithmic utility functions to represent time-tolerant applications. As in \cite{9} and \cite{16}, we can express the logarithmic utility functions as below:
\begin{equation}\label{eqn:log}
U_i(r_i)=\frac{\log(1+k_i  r_i)}{\log(1+k_i  r_{max})}
\end{equation}
where $r_{max}$ represents a user utility function that is equal to 100\%, and $k_i$ represents the increase in the rate $r_i$ in the utility percentage. Logarithmic utility functions satisfy the condition $U_i(0)=0$ and $U_i (r_{max})= 1$. Our utility proportional fairness function, as in \cite{10}, is given by
\begin{equation}
\underset{\textbf{r}}\max  \prod_{i=1}^{M} U_i(r_i) 
\end{equation}   
where $M$ is the number of users, and $r$ equals the rate of the users from 1 to $M$. The objective of our utility proportional fairness function is to maximize the utility functions of all the UEs within the system. In the mean time, it insures proportional fairness among UEs utility functions. In addition, our utility proportional fairness function prioritizes the users with real-time applications (or sigmoidal-like utility functions) while guaranteeing a minimum QoS for all the subscribers. Our main objectives are maximizing system utility and fair allocation of the resources among the users. Therefore, the optimization equation will be written as in \cite{11}:
 \begin{equation}\label{eqn:opt_prob_fairness}
\begin{aligned}
& \underset{\textbf{r}}{\text{max}}
& & \prod_{i=1}^{M}U_i(r_i) \\
& \text{subject to}
& & \sum_{i=1}^{M}r_i \leq R\\
& & &  0 \leq r_i \leq R, \;\;\;\;\; i = 1,2, ...,M.
\end{aligned}
\end{equation}
The optimization problem is a convex optimization problem and there is a tractable global optimal solution as in \cite{11}. 
\section{DISTRIBUTED OPTIMIZATION ALGORITHM }\label{sec:DISTRIBUTED OPTIMIZATION ALGORITHM}
\begin{figure}
\centering
	\includegraphics[scale=0.35]{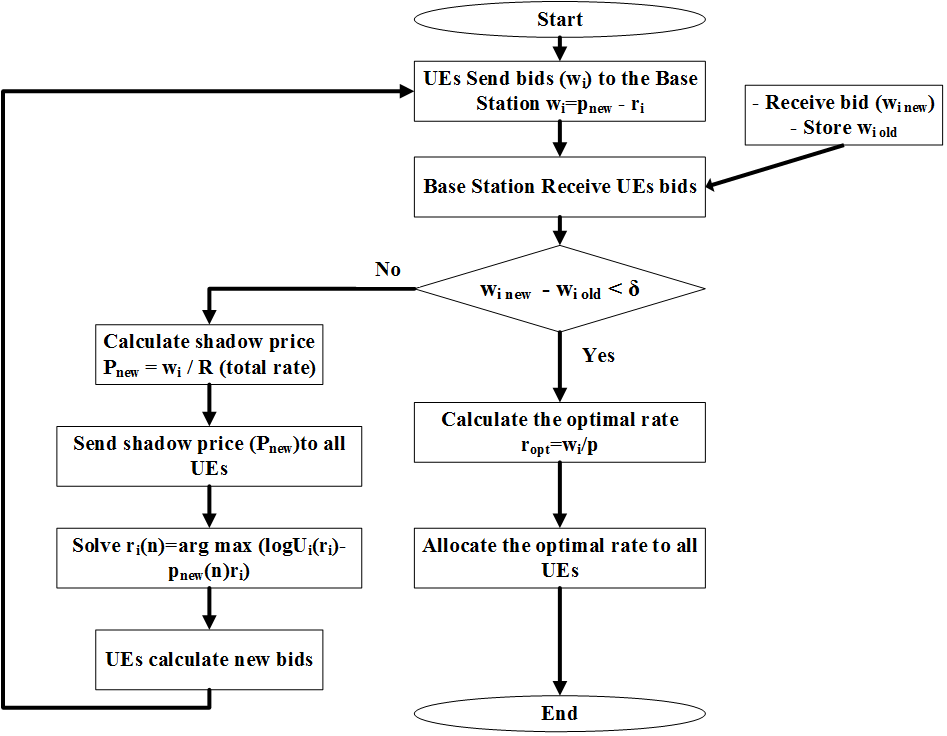} 
		\caption{Flow Chart of the Distributed Algorithm}\label{fig:FLOWCHART}
\end{figure}

Interest in multi-stage resource allocation optimization algorithms has widely increased after proving its effectiveness in solving the optimization problem in the wireless and wired networks. Our distributed optimization algorithm, which described in \figref{FLOWCHART}, is a modified version of the resource allocation optimization algorithm in \cite{1} and \cite{2}. The auction-based optimization algorithm is divided into two algorithms, at base station and UEs. 
\begin{itemize}
\item[A.] Resource allocation optimization algorithm at Base Station:\\
First of all, the base station receives bids from the UEs at time $(t)$ and then replaces the old bids at time $(t-1)$, while storing the old bids in the base station local memory. The base station compares the received bids $w_i (t)$ with $w_i (t-1)$. In case $w_i (t) - w_i (t-1) \le \delta$ (pre-specified threshold), the base station calculates the optimal rates $r_i^{opt}\big(\frac{w_i}{p})$ and send them to the UEs; otherwise, the base station computes a new shadow price $p(t)=\frac{\sum_{i=1}^{M}w_i (t)} {R}$ and communicates it to all the UEs.
\item[B.] Resource allocation optimization algorithm at the UEs:\\ 
From time to time, the users receive new shadow price $p(t)$. UEs update and replace the new $p(t)$ and store the old shadow price, $p(t-1)$. The UEs calculate the optimal rate, based on the newly received $p(t)$, and then solve the optimization problem $r_i(n) = \arg \max \big(\log U_i(r_i) -p(t) r_i )$ that subjects to the condition $r_i(n) \ge 0$, to insure that all subscribers receive a minimum QoS. The optimization problem, is solved by using bisection method and convex optimization techniques. The rate $r_i(n)$ is used to calculate the new bid $w_i(t)$ that will be sent to the base station. This process is repeated, yet the value of $w_i (t) - w_i(t-1)$, at the base station, is less than the pre-specified threshold value ($\delta$).
\end{itemize}

\section{Rate Variations}\label{sec:Rate Variations}
It is a serious challenge to maintain a constant quality of service, when user demand and available resources dynamically change. Maintaining quality service requires an appropriate allocation of resources. Adaptive real-time applications, like video and voice applications, usually require higher rates than delay-tolerant applications. In order to maintain high quality service, sufficient resources need to be allocated to the adaptive real-time applications. Allocation of resources needs to adapt to the changes in the application rates. In different adaptive real-time applications, such as video-conferencing and VoIP, the major effect on the bit-rate variations is a combination of the activity of the picture, and the picture coding method \cite{12}. In the case of video streaming, the motion of the objects, in the video, leads to variations in the bit-rate, which increases as the movement of the objects increase. Long period of no motion or few movements of small objects may result in a low bit-rate increase \cite{13}. In the past, many efforts aimed to reduce the variations in the bit-rate by means of buffering, as in \cite{12}. This increase or decrease of the bit-rate is not constant, but is dynamic and continuous. The change in the quality of video streaming due to the variations in rate affect the Quality of Experience (QoE) of the viewers, which eventually impacts customer satisfaction \cite{14}. In \cite{2}, \cite{6} and \cite{11}, the authors represent adaptive real-time application in a fixed-rate form, meaning the rate of applications does not change with time, which would never occur in reality. Our distributed algorithm considers the fluctuations of rates in the adaptive real-time applications. Assigning random distribution to the application utility functions means that the utility functions of applications continuously and randomly change with time to realistically represent the application rates.

\section{Representation of UEs Rates }\label{Sec:Representation of User Equipment Demand }
\begin{figure}
\centering
	\includegraphics[scale=0.4]{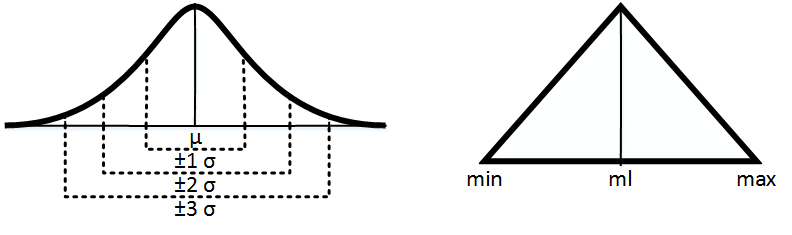}
		\caption{Normal and Triangular Distribution Functions}\label{fig:DIST}
\end{figure}

Random distributions are used to represent the randomness of variables within a certain range. In our paper, we assign two different distribution functions, normal and triangular, as shown in \figref{DIST}, to represent the continuous change in the sigmoidal-like utility functions, and then we compare the optimal rates in each case with fixed-rate applications. 

\subsection{Normal Distribution}
Normal distribution is a common representation of the random variables as the distribution is unknown. Normal distribution states the fact that approximately 97\% of the random variables are within $\pm$3 standard deviations from the mean, where $\mu$ represents mean, and $\sigma$ represents standard deviation. 

\subsection{Triangular Distribution}
In the case of triangular distribution, the values continuously and randomly vary within the $min$ and $max$ limits. The most-likely value is represented as $ml$. The applications at the $ml$ rate will work effectively. The $min$ is the minimum point where below this point the application will not function as required or stop. For example, in video applications, the video transfer stops and the image disappear. The $max$ is the maximum rate at which the application runs with the highest quality. The triangular distribution can be a good representative of a real-life situation when the minimum, maximum and most-likely values can be estimated.

\section{SIMULATION RESULTS }\label{sec:SIMULATION RESULTS }
We use Arena simulation software interfaces with Microsoft Visual Basic for Applications to solve the algorithms in equation (\ref{eqn:sigmoidal}) and (\ref{eqn:log}) for different logarithmic and sigmoidal-like utility functions. In our simulation, we consider one cellular network unit with six utility functions, to represent six UEs. Three sigmodial-like utility functions corresponding to three adaptive real-time applications. On the other hand, three logarithmic utility functions corresponding to three delay-tolerant applications. Our simulation results show convergence to the optimal solution. In the following simulations, the adaptive real-time applications will have the priority over the delay-time applications in the process of rate allocation. We set the total rate $R$=100 and number of iterations $n$=20.

\subsection{Fixed Demand of Rate}\label{fixed demand of rate}
\begin{figure}
\centering
\includegraphics[scale=0.22]{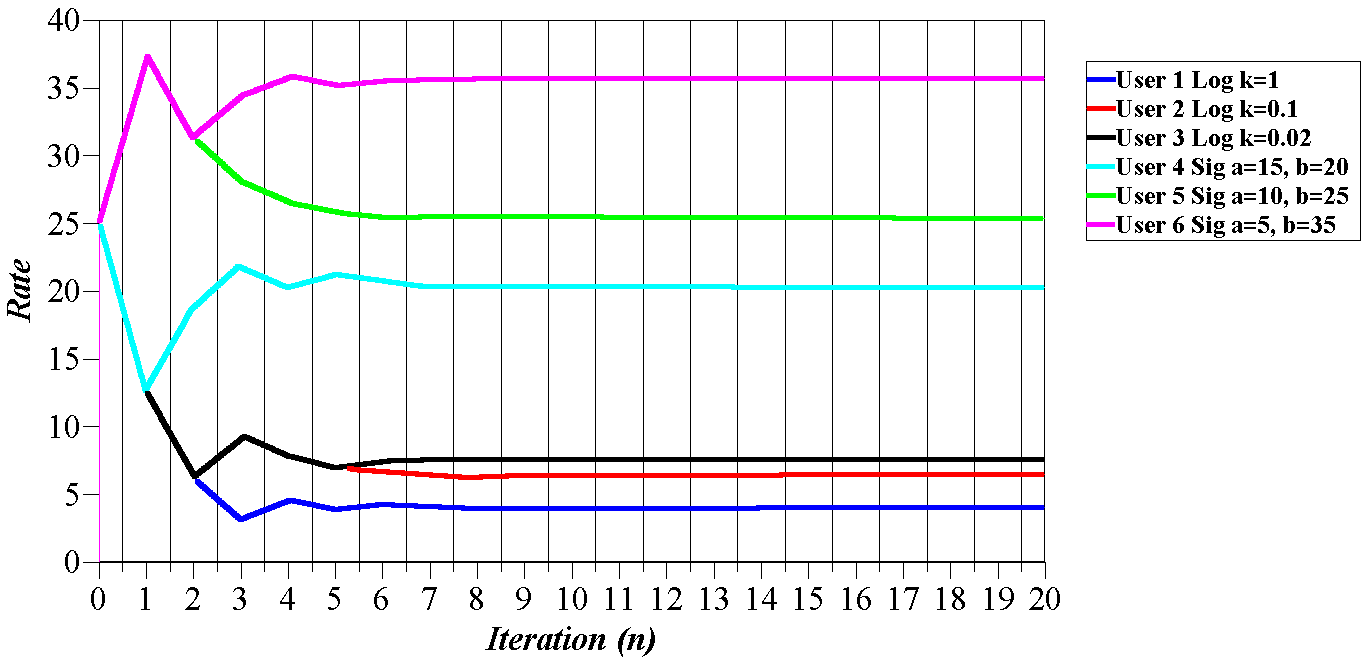}
\caption{The allocated rates to the UEs $r_i$($n$) with number of iterations $n$ when total rate $R$=100, and sigmodial-like utility function parameters are constant.}\label{fig:FIXEDRATE}
\end{figure}

\begin{figure}
\centering
\includegraphics[scale=0.22]{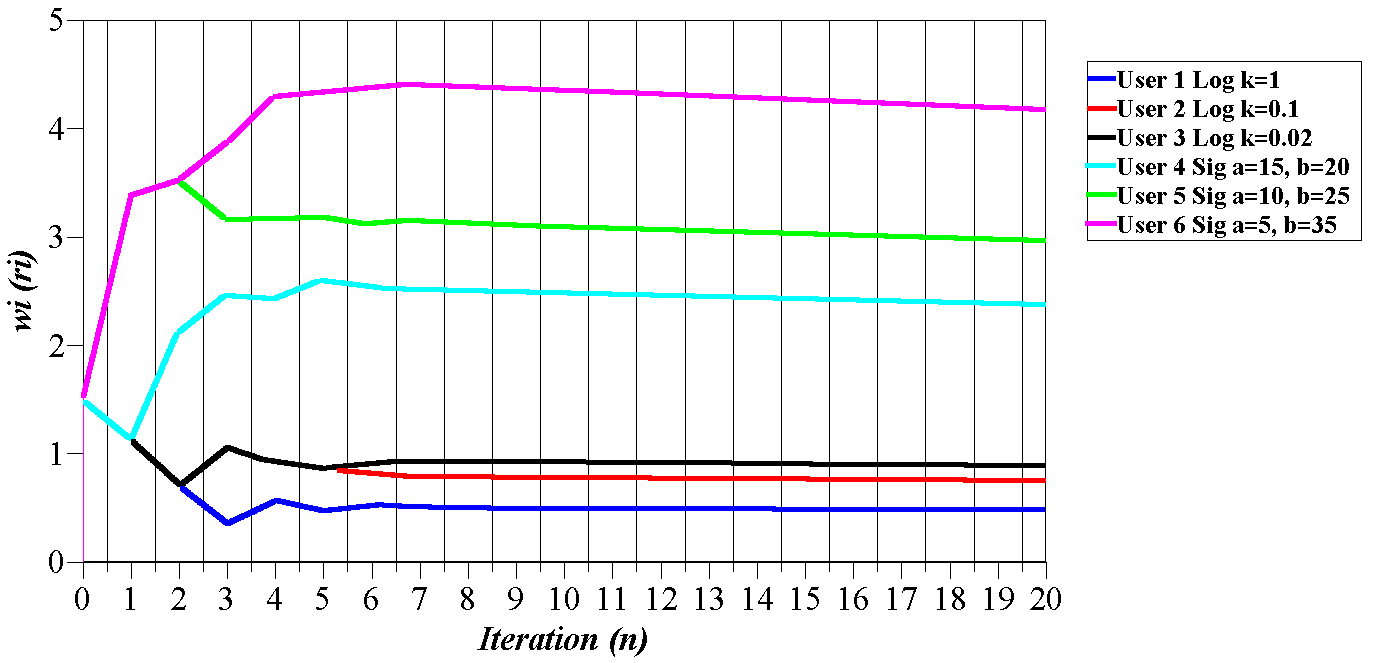}
\caption{The users bids $w_i$ ($n$) with number of iterations $n$ at total available rate $R$ = 100 as the applications rates are constant.}\label{fig:FIXEDBID}
\end{figure}

\begin{figure}
\centering
\includegraphics[scale=0.22]{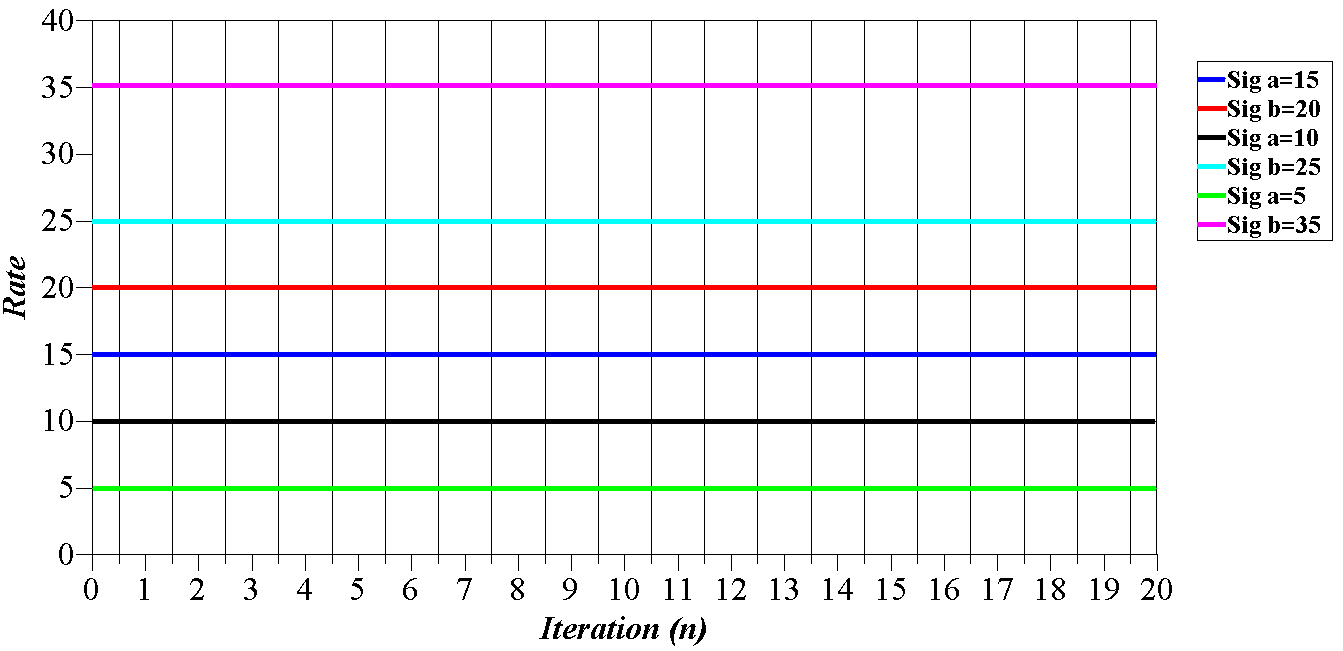}
\caption{Constant sigmodial-like utility functions parameter, $a$ and $b$, for the adaptive real-time applications.}\label{fig:FIXEDDIST}
\end{figure}

We use three logarithmic functions that are expresses in equation (\ref{eqn:log}) to represent three of the total six users. Different parameters ($k_i$) have been assigned to each of the three logarithmic users to approximately represent delay-tolerant applications (e.g. FTP). We use $k=(1,0.1,0.02)$ for users 1,2, and 3 respectively. We use three normalized sigmoidal-like functions that expressed by equation (\ref{eqn:sigmoidal}) to represent three real-time applications. Different parameters have been assigned, $a$=15, $b$=20, which is an approximation of a step function, such as VoIP, at rate $r$=20, $a$=10, $b$=25 to approximate adaptive real-time application, such as standard definition video streaming, at rate $r$=25, and $a$=5, $b$=35 to approximate adaptive real-time application, such as high definition video streaming, at rate $r$=35. \Figref{FIXEDRATE} shows the rates ($r_i$) of the users with number of iterations ($n$) in the case of fixed-rate applications. The equilibrium state of sigmoidal-like utility functions will exceed the interrelated inflection point ($b_i$). \Figref{FIXEDBID} shows the bidding process of the users with number of iterations ($n$). As the user bids increase, the higher the rate allocated. Since the rate allocation priority is given to the sigmoidal-like utility functions, the adaptive real-time applications will be given the priority of bidding and allocation of the resources, yet they reach the inflection point ($b_i$) and the remaining resources will be shared among delay-tolerant applications based on their utility parameters ($k_i$). \Figref{FIXEDDIST} shows the rates of the adaptive the real-time applications with number of iterations ($n$). The adaptive real-time application rates are fixed with time, as in \cite{2}, \cite{6}, and \cite{11}.

\subsection{Normally Distributed Demand of Rate}\label{Normal distrubtion rate}
\begin{figure}
\centering 
\includegraphics[scale=0.235]{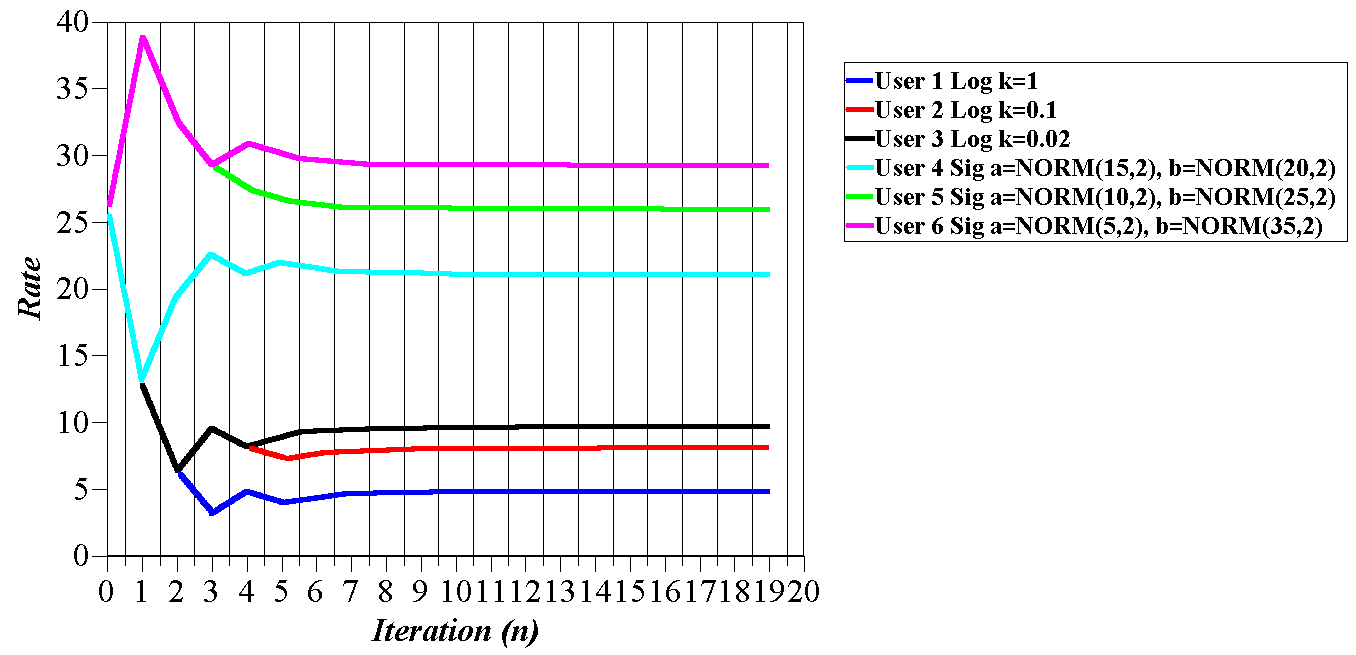}
\caption{The allocated rates convergence $r_i$ ($n$) with $n$ at total rate $R$=100. The adaptive real-time applications rates are normally distributed.}\label{fig:NORMRATE}
\end{figure}

\begin{figure}
\centering
\includegraphics[scale=0.21]{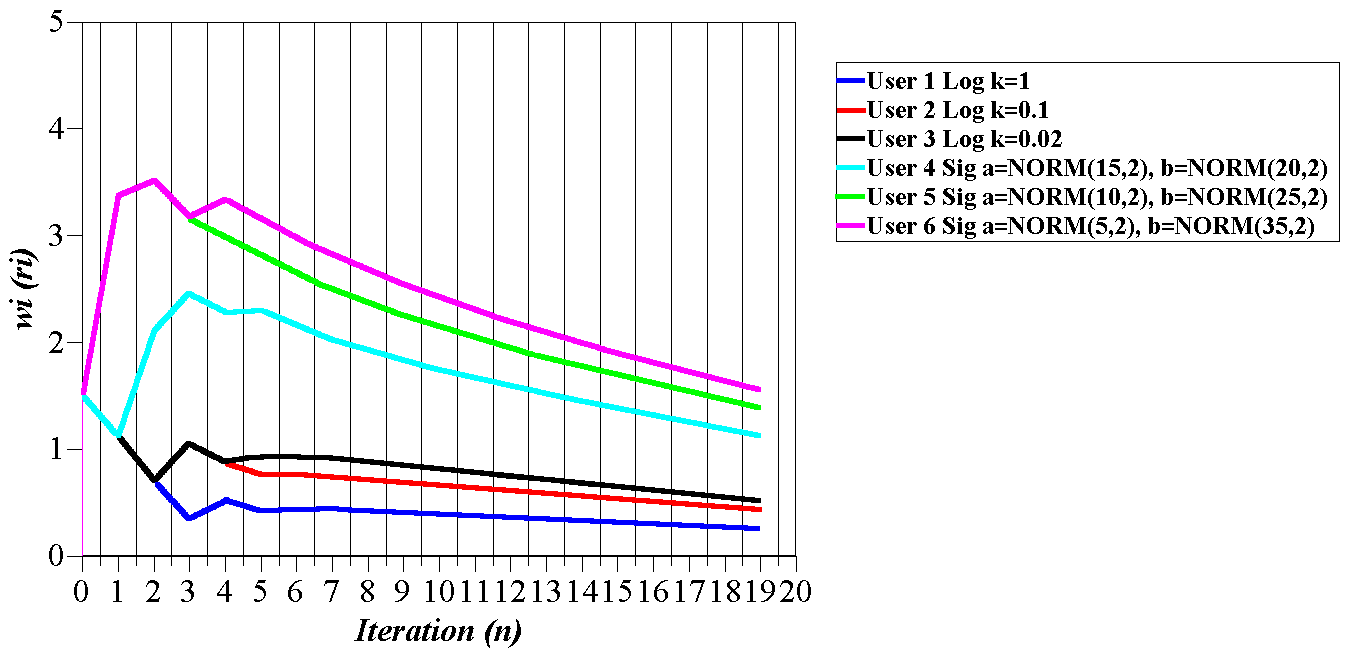}
\caption{The users bid convergence $w_i$($n$) with $n$ at total rate at the base station $R$=100. The real-time applications rates are normally distributioned.}\label{fig:NORMBID}
\end{figure}

\begin{figure}
\centering
\includegraphics[scale=0.21]{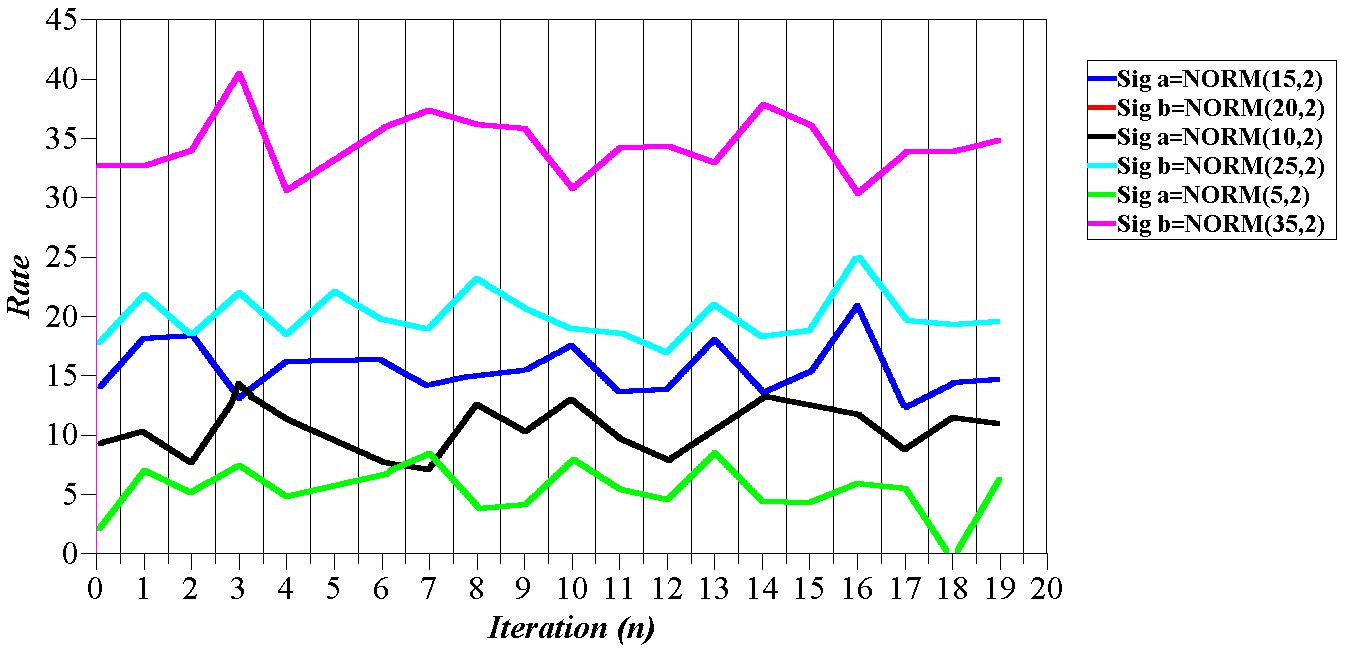}
\caption{The variations of the Sigmodial-like utility functions parameter, $a$ and $b$, for the adaptive real-time applications with number of iterations $n$.}\label{fig:NORMDIST}
\end{figure}

In the following simulations, we assign normal distribution to the sigmoidal-like utility functions parameters, $a$ and $b$. In \figref{NORMRATE}, we show the steady state rates of the UEs with with number of iterations ($n$) when we assign normal distribution to the sigmoidal-like utility functions parameters, $a=NORM(15,2)$, $b=NORM(20,2)$, represents the $4th$ user, $a=NORM(10,2)$, $b=NORM(25,2)$ represents the $5th$ user, and $a=NORM(5,2)$, $b=NORM(35,2)$ represents $6th$ user. In \figref{NORMBID}, we show the bidding process of the users with number of iterations ($n$). \Figref{NORMDIST} shows variations in the sigmoidal-like utility functions parameters, $a$ and $b$ with number of iterations ($n$).

\subsection{Triangularly Distributed Demand of Rate}
\begin{figure}
\centering
\includegraphics[scale=0.21]{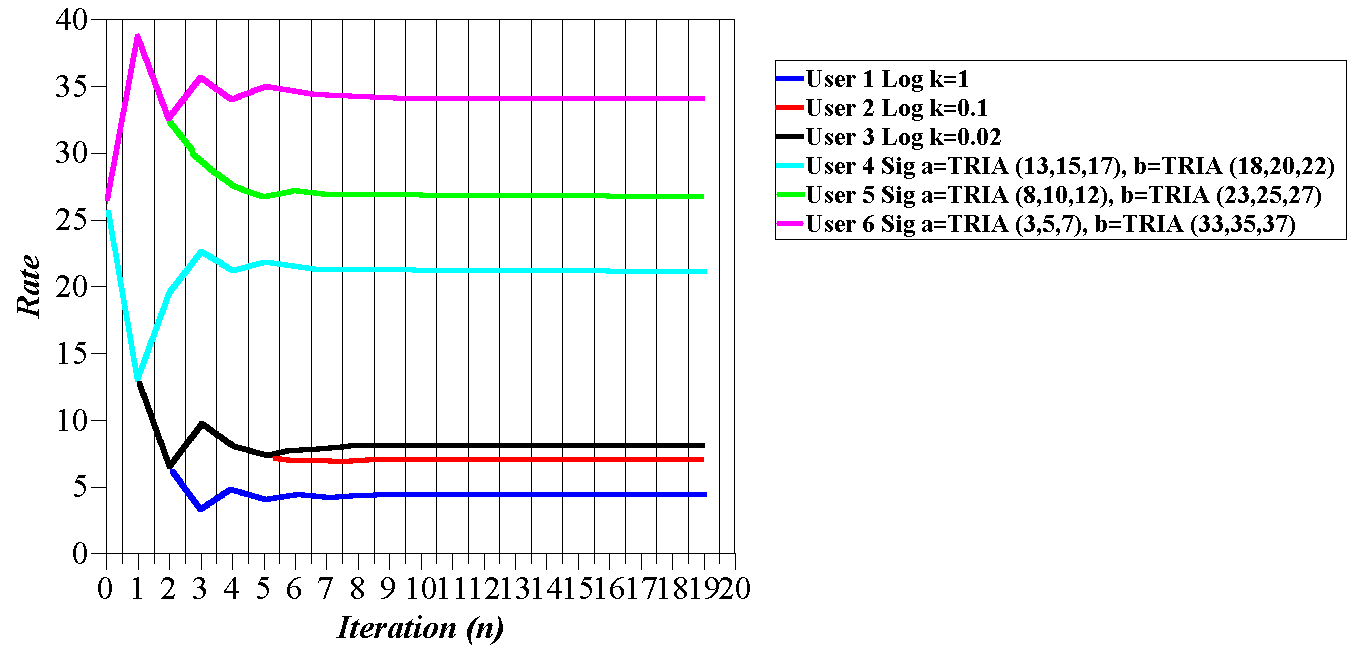}
\caption{The allocated rates convergence $r_i$ ($n$) with $n$ at total rate $R$=100. The real-time application rates change with time based on triangular distribution.}\label{fig:TRIARATE}
\end{figure}

\begin{figure}
\centering
\includegraphics[scale=0.21]{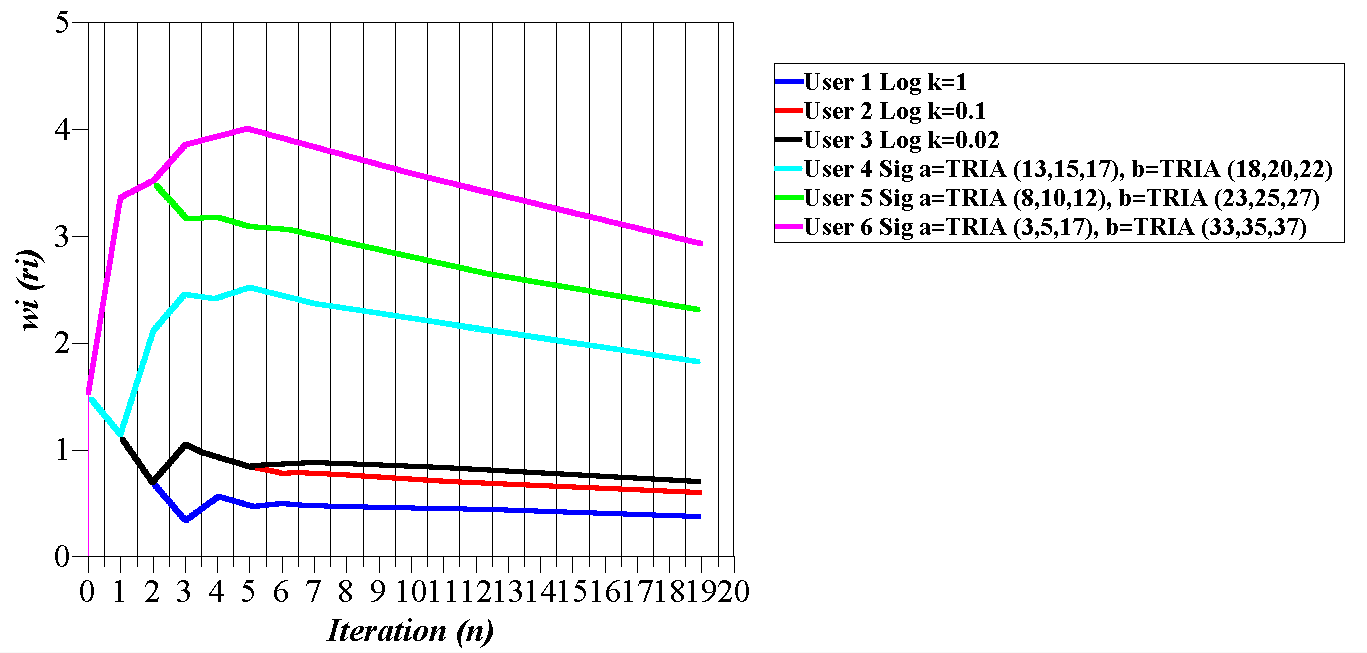}
\caption{The users bid convergence $w_i$($n$) with $n$ for total rate $R$=100. A triangular distribution assinged to real-time applications rates.}\label{fig:TRIABID}
\end{figure}

\begin{figure}
\centering
\includegraphics[scale=0.21]{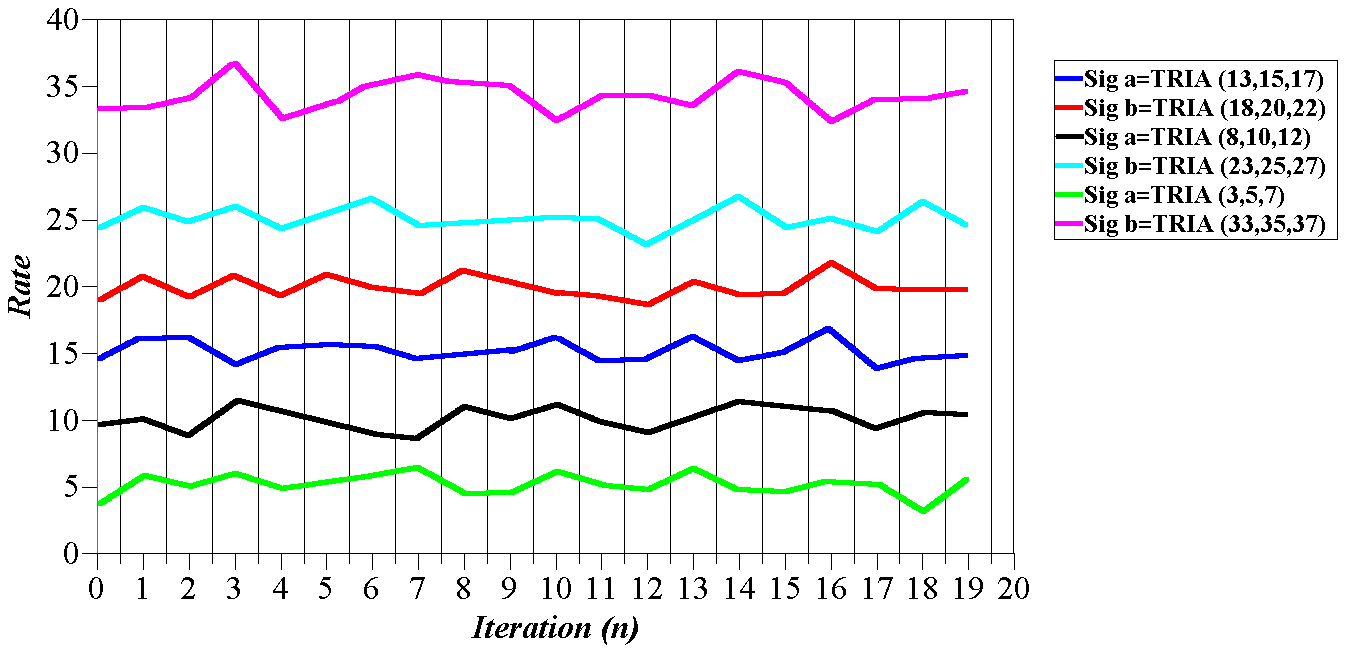}
\caption{The variation of the sigmodial-like utility functions parameter, $a$ and $b$, based on the assigned triangular distribution, with number of iterations $n$.}\label{fig:TRIADIST}
\end{figure}

In the following simulations, we assign a triangular distribution to the adaptive real-time applications. \Figref{TRIARATE} shows the optimal rate at UEs with different iterations ($n$). The sigmoidal-like utility functions parameters continuously change in correspondence to the assigned triangular distribution, $a=TRIA (13,15,17)$, $b=TRIA (18,20,22)$ which approximates the step function, and represent the $4th$ user, $a=TRIA (8,10,12)$, $b=TRIA (23,25,27)$ represents the $5th$ user, and $a=TRIA (3,5,7)$, $b=TRIA (33,35,37)$ represents $6th$ user. In \figref{TRIABID}, we show the variations in the bidding process as the sigmodial-like utility functions parameters, $a$ and $b$, vary continuously within the $min$ and $max$ limits. In \figref{TRIADIST}, we show the variations in the sigmoidal-like utility functions parameters, $a$ and $b$, based on the triangular distribution with number of iterations ($n$).
 
\section{CONCLUSION }\label{sec:CONCLUSION }
In this paper, we use a utility proportional fairness optimization model for the logarithmic and sigmoidal-like utility function applications in a cellular network. We present a distributed algorithm for optimal allocation of the base station resources to the UEs. Bisection method in addition to convex optimization techniques has been used to obtain the optimal rate allocation to the UEs. The sigmoidal-like utility functions represented in stochastic form, instead of deterministic, to provide a better optimization of resources in actual situations. Our algorithm ensures fairness in the allocation of the resources among applications. As a result, the algorithm priorities the adaptive real-time applications over the delay-tolerant applications. In addition, our distributed algorithm guarantees a minimum allocation of the resources to the users with time-tolerant applications to achieve a minimum QoS for all the subscribers. Our simulation results show that our distributed algorithm considers the variations in the rates of the real-time applications and converges to the optimal rates, while prioritizing the real-time applications over delay-tolerant applications.

\end{document}